\documentclass[preprint,review,11pt]{elsarticle}
\usepackage{graphics}
\usepackage{graphicx}
\usepackage{epsfig}
\usepackage{amssymb}
\usepackage{amsthm}
\usepackage{lineno}
\usepackage{setspace}
\biboptions{numbers,sort&compress}

\journal{International communication in heat and mass transfer}

\begin{document}
\begin{frontmatter}

\title{Effects of low-dimensional material channels on energy consumption of Nano-devices}

\author[1]{Zahra Shomali}
\author[1,2]{Reza Asgari}
\address[1]{School of Physics, Institute for Research in Fundamental Sciences (IPM), Tehran 19395-5531, Iran}
\address[2]{School of Nano Science, Institute for Research in Fundamental Sciences (IPM), Tehran 19395-5531, Iran}

\begin{abstract}
It is commonly believed that the significant energy saving advantages are belonged to the logic circuits which operate at low temperature as less enegy is needed for cooling them to the treshold temperature after operation. Also, nanoscale thermal management, efficient energy usage in nanoscale and especially thermal optimization are the most challenging issues, while dealing with the new generation of transistors as the miniaturizing unlimitedly the silicon channels of the transistors has resulted in an increase in the energy consumption of computers and the leakage currents. In this paper, the non-Fourier thermal attitudes of well-known two-dimensional crystalline materials of graphene, blue phosphorene, germanene, silicene and MoS$_2$ as the silicon channels replacements are studied by using the phonon Monte-Carlo method. We show that graphene and blue phosphorene have the least maximum temperature, representer of the reliability of the transistors, among the all five investigated nano-channels during the Monte-Cralo simulation. The established hotspots of these two materials are always cooler, not reaching the temperature threshold level, and they lose the heat much faster as the heat generation zone is switched off. The obtained results considered along with the electrical disadvantages of the graphene layer, suggests the blue phosphorene as the more thermally appropriate and optimal choice for the silicon channel replacement in new designed field effect transistors. That is to say that the limit of the energy and economic cost of the producing the advanced blue phosphorene chips meets the value of the product for the computing enterprise.
\end{abstract}

\end{frontmatter}

\section{Introduction}

Since the first silicon transistor has been launched, in the quest for higher performance, the size of these nano-devices has been incredibly decreased. Nowadays, the number of transistors on a single chip has been grown from a few thousand in the earliest integrated circuits to more than two billions \cite{Huang2017}. In spite of the recent progress in achieving the much smaller silicon field effect transistors (FETs), the size reduction of the FETs consisting of three-dimensional (3D) semiconductors is limited by the diminution of the self-generated heat dissipation rate caused by the increase in static power and the leakage current between the source and the drain. The solution for the high energy consumption and the power devaluation have been investigated to decide the transistor contradiction of the smaller size/larger energy loss. As a consequence, FETs with channels made from two-dimensional (2D) semiconductors with almost eliminated leakage current owing to the all electrons in atomically thin channels and, accordingly uniformly influenced by the gate voltage have drawn attention \cite{ghazanfarian2012}. Therefore, new transistor technology based on improving thermal efficiency by the usage of the channel candidates, chosen from the low-dimensional substances with odd thermal and electrical properties, such as graphene \cite{Schwierz2010}, germanene \cite{Eugenia2016}, silicene \cite{Lay2015,Tao2015}, phosphorene allotropes \cite{Li2014,Zhang2017,Zare2017} and Molibdeniyum disulfid (MoS$_{2}$) \cite{Radisavljevic2011}, has been activated. Each proposed low-dimensional transistor has advantages and disadvantages over the others. Therefore, the main goal of nano-electronic technology is to find the most thermally and electrically optimal nano-devices that are best suited for the substitution of the old-fashioned silicon channels and simultaneously low cost economically in cooling processes.

A decade ago, graphene which firstly attracted the attention to lower-dimensional systems, was suggested as the pioneering candidate for the silicon replacement in semiconductor device fabrication technology. Initially, the odd properties and high electrical and thermal conductivity of this appealing material made it to be announced as the perfect nominee for the transistor nanotechnology. But more studies revealed the zero bandgap of graphene which leads to the low on-off current ratios of 2 to 20. This deficiency alone, enforces researchers to behave more prudent with graphene and investigate for other matters. The current ratio, on the other hand, can be increased by shrinking the graphene sheet to one-dimension and manufacturing the graphene nanoribbons (GNR). Although, there are some proposals for using these nanoribbons as a silicon substitution \cite{Son2013}, there are serious limitations for implementing the GNRs in current nanotechnology such as the reduction of the mobility from 200,000 cm$^2$/V.S to 1000 cm$^2$/V.S and difficulty in patterning the graphene in the form of GNR \cite{Berger2006,Sprinkle2010}. In recent years, the FET community has shifted its interest to the sheets of advanced 2D crystalline materials which do not need to be altered to open their bandgap and have natural bandgaps of size more than 0.3 eV such that their low off-state currents and high on-state currents can be figured out. Particularly, phosphorene, silicene, germanene and mono- and few-layered transition-metal dichalcogenide systems (TMDs) such as MoS$_{2}$, have been proposed for FETs \cite{Chhowalla2016}.

The most intriguing candidate is phosphorus, which is grouped into the graphene category. Currently, a high interest has been raised to use phosphorene as a new 2D crystalline material for electronic applications which is mainly due to its uniqueness among all 2D substances by having both an intrinsic and sizable bandgap (unlike graphene) and a high carrier mobility (unlike most transition-metal dichalcogenides). Phosphorene has different allotropes of $\alpha$-, $\beta$-, $\gamma$-, $\delta$- and $\zeta$-phases. Although high performance FET using $\alpha$-phosphorus, known as black phosphorus has been reported recently \cite{Li2014}, the anisotropic thermal transport properties of black phosphorene may degrade the device reliability and performance since the low thermal conductivity along the armchair direction can lead the localized Joule heating in the confined system. Among the five mentioned phosphorene allotropes, it is very hopeful to use $\beta$-phase to solve the thermal management issues in the black phosphorene based FETs. Also, the interfacial thermal resistance for $\beta$-phosphorene, called as the blue phosphorene, is only a quarter to that of the graphene and this substance is more sensitive to the environmental variations than silicene \cite{2Zhang2017}. In addition, a similar lattice structure and thermal conductivity to MoS$_{2}$ monolayer will make blue phosphorene a promising material to form the MOSFET channel, whose electronic properties is highly tunable by controlling the layer thickness and stacking order \cite{Zhang2017}.

The other nominees for the replacement of the silicon in the transistor industry could, in fact, be silicon itself in the form of silicene, or its sister material,  germanene. Silicene which is the silicon-based analogue of graphene, owing to its buckled structure, favors the existence of a bandgap of 0.21 eV and consequently the on/off ratio of 10. A parent silicon crystal, like what graphite is for graphene, does not exist in nature, but silicene is synthesized using the scalable epitaxy method. The measured mobility for both electrons and holes is about 100 cm$^{2}$/V.S, which is quite low compared with that of graphene and blue phosphorene \cite{Lay2015,Tao2015} and somehow larger than the carrier mobility in the single-layer MoS$_{2}$. If we look more closely to the single layer of molybdenum disulfide, it has a large intrinsic bandgap of 1.8 eV \cite{Mak2010} and the mobility of 0.5-3 cm$^2$/V.S, which is too low for pragmatic transistor nano-devices. In practice, a 30 nm hafnium oxide,  as a high-k gate dielectric on the top is used for boosting the mobility upto the value of 200 cm$^2$/V.S. As a consequence, although, some works suggest these nano materials as the promising candidates for 2D transistors, the silicene with its relative low carrier mobility and MoS$_{2}$ with its costs for making use of the full potential of it, seems not to be suitable alone for the new developed field effect transistors. On the other hand, for the formation of a germanene based field-effect transistor, a band gap opening of at least 0.3-0.4 meV is required. The band gap opens in germanene by applying an external electric field in a direction perpendicular to the germanene \cite{Chen2016}. In similarity to the graphene, the band gap also opens by shrinking the material to nano-ribbon. Accordingly, the experimental challenging existed in graphene nanoribbons fabrication also appears for germanene nanoribbons and therefore this solution for band gap opening is not the adequate one. On the other hand, the large intrinsic carrier mobility of germanene which is obtained to be a factor of 2-3 higher than the intrinsic carrier mobilities of silicene is the advantage of this material. However, germanene is not as compatible with the current silicon-based Microtechnology as silicene \cite{Acun2015}.

In all cases of energy application, more than 90$\%$ of primary energy is first converted to heat and all attempts are trying to reduce this huge amount of dissipation. This is also the important consideration while dealing with the transistors. Concerning all these new candidates of FET technology, the most important common issue is the self-heating procedure \cite{Sulima2007}. The continued miniaturization of integrated circuits and the current trend toward nanoscale electronics have led to tremendous integration levels, with hundreds of millions of transistors assembled on a chip area no larger than a few square centimeters. Circuit densities are projected to reach the giga-scale as the smallest lateral device feature sizes approach 10 nm. The bottleneck technological issue of this scaling trend is the power problem. That is to say that the chip temperatures become too large due to the self-heating and the power densities such that it prevents the reliable operation of the integrated circuits. The chip-level power densities which are currently on the order of 100 W/cm$^2$ will increase much as the more integration and shrinking occurs due to the International Technology Roadmap for Semiconductors (ITRS) guidelines. This high power density and self-heating makes the batteries to drain briskly in portable devices and accordingly they must be cooled to a temperature at which the transistors can operate. Consequently, as long as the cooling technology has not got up with the miniaturization many electronic devices are impractical due to the millimeter-scale hotspots on the chip, which are localized zones with higher heat per unit area and accordingly higher temperatures \cite{Balachandra1991,Pop2006-2,Pop2010,Samian2013,Samian2014,Gong2015}. On the other hand, while chip-level hotspots are troubling circuit designers \cite{Mahajan2002,Pop2005}, device designers are beginning to encounter thermal management problems at nanometer-length scales within individual transistors. Also, novel and complex device geometries tend to make heat removal more difficult and most new 2D materials being introduced in device processing have lower thermal conductivities than bulk silicon. Concerning FET, the applied voltage establishes a lateral electric field with a peak near the drain. This field accelerates the charge carriers and hence they can scatter from the other electrons, phonons, material interfaces and imperfections. The electron population gives out the energy by scattering from phonons while self-heating the lattice through the well-known Joule heating mechanism. Therefore, the Joule heating in FETs results in a local temperature rise (hot spot) where local temperatures are remarkably higher than the die average temperature. It has been obtained that among the all heat generation effects such as Joule heating, current crowding, thermoelectric, the Joule heating is evidently the dominant self-heating for low-dimensional systems \cite{Grosse2011}.

In nano-electronics technology, the overall reliability is rectified by the temperature of the hottest region on the die, instead of the average die temperature. Thus, hotspots often control the required higher-level packaging and thermal management solutions including the material selections and the heat spreader design \cite{Moore2014}. As a consequence, the thermal behavior of the nano-device and the die has a decisive role in finding the cooling requirements and controlling the impact of the environment. In other words, as the power density increases due to the growth of the temperature, the performance of the nano-device may be throttled to cause the power to decrease. Therefore, many efforts have been performed to enhance and optimize the cooling techniques for the increase of heat spreading at both the single transistor nano-device and the global die as well as the chip for minimizing the intensity and impact of the hotspots \cite{Shomali2016}. Apart from the attempts for achieving the desired methods of heat spreading, among the low-dimensional silicon replacement nominees, the one with the minimum peak temperature rise is the most optimal and appropriate choice for nano-electronics industry. This is true as the transistor cooling to keep the hottest zones of the dice under stated temperature threshold is more feasible for nano-devices with lower maximum temperatures. Indeed, the operation of the new generation transistor technology used in new modern digital devices such as laptops, tablets, and cellphones completely depends on the thermal attitude of these nano-devices. Thermally inefficient working of the transistors results in unlimited increasing temperature which can make the digital device stop working permanently. The importance of energy efficiency in nano-scale for better cooling and operation of the transistors is such that there are many works devoted to this field \cite{Pop2010}.

In this paper, a numerical framework to evaluate the extremely low-power and most energy saving FETs using heterostructures of 2D semiconductors is provided. To be more precise, the formation of the hot spot in the presence of the self-heating is investigated for five major candidates of silicon channel replacements. The Monte-Carlo (MC) simulation is used to study the transient heat conduction in graphene, blue phosphorene, germanene, silicene, and MoS$_{2}$ channel transistors. Having mentioned before, for nano-electronics the overall reliability and thermal efficiency are determined by the hottest region of the transistor. Hence, applying the heat generation zone of homogeneous heat flux, the low-dimensional transistor with the most adequate conditions and lowest maximum temperature is discovered. This paper is organized as follows. In Sec. \ref{geo} of this paper, the geometry and the boundary conditions are discussed. Then in Sec. \ref{Sec.3}, the mathematical modeling is given. Sections. \ref{results} and \ref{con}, subsequently, demonstrate the results and the conclusions.

\section{Geometry and boundary conditions}
\label{geo}

\begin{figure}
\vspace*{-7.0cm}
\hspace*{0.4cm}
\centering
\includegraphics[width=0.7\columnwidth]{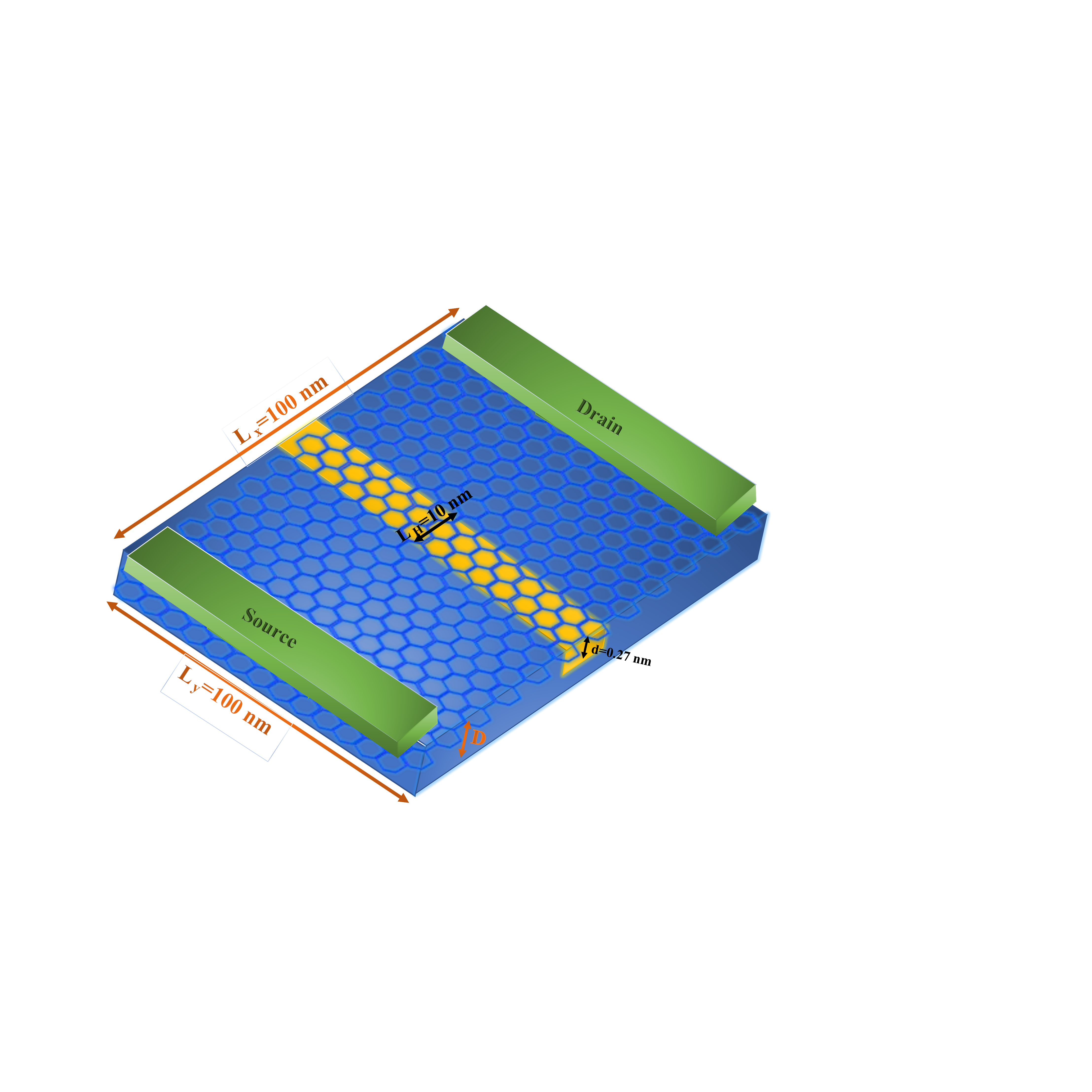}
\vspace*{-2.8cm}
\caption{\label{scheme} The low-dimensional material MOSFET, as an example graphene FET is shown here, with all boundaries adiabatic except the bottom boundary which exchange the energy with the surroundings. L$_x$, L$_y$, d, and D are, respectively, the length along the $x$ direction, the $y$-length, the thickness of the layer, and the assumed depth of the heat generation zone.}
\end{figure}

A single layer large area of five important low-dimensional material sheet as the MOSFET transistor channel is simulated by the usage of the non-Fourier atomistic Monte-Carlo simulation. The self-heating source and the boundary conditions are almost identical to what exists in a real FET nano-device. The sizes of the simulated transistors are L$_{x}$=100 nm, L$_{y}$=100 nm, with the thickness of 0.35, 0.53, 0.40, 0.27, and 0.65 nm, respectively, for graphene, blue phosphorene, silicene, germanene, and MoS$_{2}$.  As seen in Fig. \ref{scheme}, the heat generation volume of 10$\times$100$\times$0.27 nm,  is contemplated at the center of the studied layers, starting at L$_x$-L$_H$/2 with L$_H$ being the width of the heat generation zone. The depth d=0.27 nm, is the thickness of the germanene, the thinnest analyzed low-dimensional substance. The value of the self-heating source is assumed to be Q=2.5$\times$10$^{12}$ W/m$^{3}$. All boundaries of the investigated new low-dimensional materials are assumed to be adiabatic except the bottom boundaries which can exchange the energy with the environment \cite{Shomali2017,2Shomali2017}. Consequently, the temperature at these boundaries are set to the ambient temperature. The initial temperature of the all segments of the devices is kept at 299 K. The most part of the generated heat inside the transistor is dissipated to the surroundings and is conducted toward the bottom. Also, the flat-edge low-dimensional material (rectangularly shaped) with specular reflection in our simulation are studied. Therefore, as specular reflection from flat edges does not alter the phonon velocity component along the substance, it introduces no resistance.

\section{Mathematical modeling}
\label{Sec.3}

Phonons as the main carriers of the heat in most low-dimensional materials can be treated as the semiclassical particles. Accordingly, the Boltzmann transport equation is used to govern phonons via describing the phonon distribution function, f$_b$$(\mathbf{r},\mathbf{q},t)$ depending on the location, $\mathbf{r}$, wave vector, $\mathbf{q}$ and time t for each phonon branch, b. The phonon Boltzmann transport equation including both equilibrium and non-equilibrium phenomena is written as $\frac{\partial \textrm{f}_{b}(\mathbf{r},\mathbf{q},t)}{\partial t}+\textrm{v}_{b,\mathbf{q}} \ . \ \nabla_{r}\textrm{f}_{b}(\mathbf{r},\mathbf{q},t)=\frac{\partial \textrm{f}_b (\mathbf{r},\mathbf{q},t)}{\partial t}\mid_{scat}$. Here, v$_{b,\mathbf{q}}$ is the group velocity of phonons where $\omega$$_{b,\mathbf{q}}$ is the angular frequency of branch $b$ with the wave vector of $\mathbf{q}$. In general, the frequency depends on both the value and the direction of the wave-vector. In isotropic assumption the frequency only relies on the value of the wave-vector and is the same in all directions of the space. In this paper, large-area sheets of graphene, blue phophorene, germanene, silicene, and MoS$_{2}$ are investigated. Hence, like the silicon and the other isotropic materials the phonon dispersion relation of these sheets will be considered isotropic. It is expected that the resulted error is small owing to the fact that the lattice thermal conductivity of a large area of the desired low-dimensional materials with the names of graphene, blue phosphorene, silicene, germanene and MoS$_{2}$ is isotropic \cite{Lindsay2010,Wei2014,Jain2015,Liu2015,Zhang2017}. Also the heat transport in wide width nanoribbons is spatially isotropic \cite{Aksamija2012}. Here, we use the stochastic phonon Monte-Carlo (PMC) method to solve the PBTE. Modelling with PMC starts with considering the equilibrium conditions. The phonon distribution function at absolute temperature $T$, in the equilibrium situation is the famous Bose-Einstein (BE) distribution. In the MC method, phonons considered as statistical samples are assigned to six individual stochastic spaces: three wave vector and three position vector components. The sampled phonons firstly experience the drift (ballistic motion) and then take part into the scattering processes. The steps in an MC simulation of the low-dimensional material channel sheets are briefly expressed in the following. First, the number of phonons per unit volume, ubiquitous all over the computational domain, $N_{\rm actual}$, is calculated via the relation $N_{actual}=\sum_{p} \int[\exp(\frac{\hbar \omega}{k_{B}T})-1]^{-1}D(\omega_{o,i})d\omega.$ Here, D$(\omega_{o,i})$=$\frac{1}{2\pi^{2}} |\frac{d|\mathbf{K}|}{d\omega}|$ is the phonon density of states. The integration is over the whole frequency space for each phonon polarization which is numerically performed by discretizing the frequency space. The maximum frequency in each branch, $\omega_{max}$, is driven from each dispersion curve. The frequency interval between $0$ and $\omega_{max}$ are divided into the $N_{int}$ intervals with $\omega_{o,i}$ being the central frequency and the bandwidth of the $i$-th spectral bin, respectively. This reduces the integral into a summation over discrete frequency intervals. In this paper, $\Delta\omega_{i}$ is taken to be 2.54$\times$10$^{11}$, 3.93$\times$10$^{10}$, 1.61$\times$10$^{10}$, 2.63$\times$10$^{10}$, 4.5$\times$10$^{10}$, respectively, for graphene, blue phosphorene, germanene, silicene and MoS$_{2}$ and N$_{\rm int}$ is 1000. It is shown that for the studied sheets, these values of the parameters guarantee the accurateness of the solutions. The actual number of the phonons, N$_{\rm actual}$, is usually a very large number which makes the modeling to be computationally expensive. Hence, by introducing the scaling or a weighting factor, W=$\frac{\textrm{N}_{\rm actual}}{\textrm{N}_{\rm prescribed}}$, the number of phonons which are simulated, are decreased. N$_{\rm prescribed}$ is the number of phonons (stochastic samples) actually initialized/emitted into the system. In other words, each stochastic sample is the representative of an ensemble of W phonons which should be initialized. The simulation domain is a sheet of width W and length L with the thickness of D. The sample phonons are spatially initialized by dividing the computational domain into a number of cells or control volumes. Three random numbers of R$_{1}$, R$_{2}$ and R$_{3}$ are drawn to locate each phonon as $\mathbf{r}$=R$_{1}$W$\mathbf{\hat{x}}$+R$_{2}$L$\mathbf{\hat{y}}$+R$_{3}$D$\mathbf{\hat{z}}$. With due attention to the number of phonons in the $i$-th spectral frequency bin, a normalized frequency cumulative density function, CDF is constructed as $\textrm{F}_{i}=\frac{\sum_{j=1}^{i}\textrm{N}_{j}}{\sum_{j=1}^{N_b}\textrm{N}_{j}}$ with $F_{i}$ representing the probability of finding a phonon with frequency less than $\omega_{o,i}$+$\Delta\omega_{i}$. Drawing a random number R$_{F}$, when F$_{i-1}$$<$R$_{ F}$$<$F$_{i}$ is satisfied, the $\omega$ will be calculated as $\omega_{o,i}$+(2R$_{F}$-1)$\frac{\Delta \omega_{i}}{2}$. The probability of belonging the frequency to the polarization, b, in the $i$-th spectral frequency bin is given by, $\textrm{P}_{i,p}=\frac{\textrm{N}_{i,b}}{\textrm{N}_{i}}+\textrm{P}_{i,p-1}$ where N$_{i,b}$ is the number of phonons in branch b and N$_{i}$ is the total number of phonons in the $i$-th spectral frequency bin. Then, a random number, R$_{P}$ is taken into account. If P$_{i-1}$$<$R$_{\rm b}$$<$P$_{i}$, then the phonon is ascribed to the polarization $P$ or branch b. In the present study, only three transverse (TA), longitudinal (LA), and flexural (ZA) for each low-dimensional material are taken into account. Once the frequency and polarization of the phonon are established, the value of its group velocity is calculated via v$_g$=$\nabla$$_{k}$$\omega$. In the last step of initialization, the direction is attributed to each phonon sample (for more details of initialization see \cite{2Shomali2017}).

The intransitive and important parameters for PMC simulation are the phonon dispersion curves which demonstrate the relation between the frequency $\omega$ and the wave vector $\vec{q}$ in the first Brillouin Zone (1BZ). These curves reflect the symmetry of the underlying lattice. From dispersion, the properties such as group velocity and density of states can be extracted and used in the solution of the BTE. Graphene, blue phosphorene, germanene, silicene, and MoS$_{2}$ has six phonon branches. Importantly, the occupation of optical phonon states is low at temperatures of up to 600 K. Hence, as many studies confirm, the contribution from optical branches is small and negligible in heat conduction and only acoustic branches should be taken into account \cite{Mei2014}. From the symmetry analysis of crystals, it is known that the systems are thermally isotropic when the order of the symmetry is higher than C$_6$. Here, the studied low-dimensional systems are all hexagonal lattice systems with C$_6$ symmetry \cite{Liu2015}. Consequently, the quadratic fitting to the full dispersion relations of the all five low-dimensional materials \cite{Ge2016,Nasr2017}, is performed to obtain an approximate isotropic relation. Consequently, $\omega$$_b$=c$_{b}$k$^{2}$+v$_{b}$k is acquired for each low-dimensional material. All the obtained coefficients are provided in Table \ref{Tab1-tab1}.
\begin{table*}[htbp]
\caption{The c$_{LA,TA,ZA}$ and v$_{LA,TA,ZA}$ coefficients of the fitted quadratic formula of $\omega$$_b$=c$_b$k$^{2}$+v$_b$k for five studied low-dimensional materials. \newline}
 \label{Tab1-tab1}
\centering
\begin{small}
\hspace*{-0.9cm}
\begin{tabular}{|c|c|c|c|c|c|}
  \hline
2D Material  &  Graphene  &
Blue phosphorene  & Germanene  & Silicene  & MoS$_{2}$   \\ \hline
c$_{LA}$(m$^2$/s)  & -3.91$\times$10$^{-7}$  & -5.9$\times$10$^{-7}$   & -5.20$\times$10$^{-7}$  & -7.45$\times$10$^{-7}$  &  -3.35$\times$10$^{-7}$ \\ \hline
c$_{TA}$(m$^2$/s) & -4.85$\times$10$^{-7}$ & -3.5$\times$10$^{-7}$ & -2.31$\times$10$^{-7} $ & -4.14$\times$10$^{-7}$ & -1.98$\times$10$^{-7}$  \\ \hline
c$_{ZA}$(m$^2$/s) & 1.92$\times$10$^{-7}$ & 1.61$\times$10$^{-7}$  & 0.992$\times$10$^{-7}$ & 1.64$\times$10$^{-7}$ & 0.467$\times$10$^{-7}$  \\ \hline
v$_{LA}$(m/s) & 23050 & 9668 & 5787 & 8848 & 7772 \\ \hline
v$_{TA}$(m/s) & 15294 & 5897 & 3179 & 5915 & 4880  \\ \hline
v$_{ZA}$(m/s) & 2670 & 896 & 329 & 677 & 2522 \\ \hline
\end{tabular}
\end{small}
\end{table*}

At the next step, the initialized phonons, are allowed to move from one point to another. During this transport, the phonons may experience impacts with other phonons, or scattering from the boundaries. In the MC technique, the drift and the scattering events are, subsequently, treated independently. The positions of the drifted phonons inside the low-dimensional sheets are traced via using an explicit first order time integration. The phonon drift results in redistribution of the energy (and consequently the temperature) in the computational domain. At the end of each drift, the energy of the all phonons of each cell are summed and consequently the energy of each spatial frequency interval per unit volume, $\tilde{U}_{cell}$, is calculated. Then, the pseudo-temperature $\tilde{T}$, the parameter defined for simplifying the calculation of the scattering processes, is calculated via Newton-Raphson method \cite{2Shomali2017}. Also, each phonon which travels along the direction $\hat{s}$ may hit the boundaries. Just that the boundaries at which the phonons hit are determined, the behavior of the phonon ensembles after the collision is investigated. A phonon that hit the boundary can be reflected specularly or diffusively. In the present paper, low-dimensional sheets with flat boundaries is studied. Since specular reflection does not employ any resistance to the thermal transport, the phonon velocity along the boundary does not change. A phonon which travels in a solid can be scattered by means of lattice imperfections, interactions with electrons, interactions with other phonons (intrinsic scattering), and boundaries. Each of these scattering, which itself can be elastic (both energy and momentum are conserved) or (only the energy is conserved), result in exchanging energy between different lattice wave vectors. Here, as the graphene, blue phosphorene, germanene, silicene, and MoS$_{2}$ are assumed to be pure and perfect, the scattering due to vacancies, dislocations and impurities is neglected. Three-phonon (Umklapp and normal) scattering and scattering from the boundaries are the mechanisms which are considered in the present Monte-Carlo study of the low-dimensional systems. Phonon-phonon scattering is the most important mechanism in the usual temperature range of interest (300-600 K) for a suspended low-dimensional system. Three-phonon processes can be a normal (N) processes or an Umklapp (U) processes. While N processes do not directly cause the thermal resistance and only play a role through redistributing the phonons, U scatterings are the reason of the thermal resistance. The relaxation-time approximation, a simplified description of the scattering dynamics cannot perfectly describe heat transport at low temperatures \cite{Cepellotti2017}. Here, as we deal with the systems at almost high temperature, this approximation is adequately accurate. Accordingly, the three phonon interactions are treated through a scattering rate \cite{Mazumder2001}. The Umklapp phonon-phonon scattering rate is calculated from the standard general approximation for dielectric crystals \cite{Liu2015}, $\tau^{-1}_{b,U}(\omega)=\frac{\hbar \gamma ^{2}_{b}}{\bar{M} \Theta_{b} v^{2}_{s,b}} \omega^{2} T e^{-\Theta_b/3T}$. In this equation v$_{s,b}$ is the sound velocity of branch b and $\bar{M}$ is the average atomic mass. Also, $\Theta$ is the Debye temperature. The first part of the rate is the standard Umklapp interaction strength, while the exponential term calculates the effective contribution from the redistribution via the N processes. In this paper the values of $\Theta$, $\bar{M}$ and also, the Gr\"{u}neissen parameters, $\gamma$$_{TA,LA,ZA}$ are demonstrated in the table \ref{Tab2-BP}.
\begin{table*}[htbp]
\caption{The thermal parameters used for relaxation time approximation formula.  \newline}
 \label{Tab2-BP}
\centering
\begin{small}
\hspace*{-0.4cm}
\begin{tabular}{|c|c|c|c|c|c|}
  \hline
2D Material & Graphene &
Blue phosphorene  & Germanene  & Silicene  & MoS$_{2}$   \\ \hline
$\Theta$  & 1911 & 550 & 352  & 680  &  687 \\ \hline
$\bar{M}$(e$^{-27}$ kg) & 19.94 & 51.4102 & 120.6215 & 46.64457 & 159.3121  \\ \hline
$\gamma$$_{LA}$ & 2.0 & 2.17  & 1.11 & 2.17 & 1.142  \\ \hline
$\gamma$$_{TA}$ & 0.67 & 1.988  & -0.4 & 2.2 & 0.920 \\ \hline
$\gamma$$_{ZA}$ & -1.53 & -1.03  & -3.5 & -5.03 & -2.867  \\ \hline
\end{tabular}
\end{small}
\end{table*}
The phonon-phonon scattering with the presence of four or more phonons are not taken into account. This is due to the fact that this type of scattering gets important at temperatures much higher than the Debye temperature which is far from the operating temperature of most nano-electronic devices. In Monte-Carlo simulation, the probability of scattering a phonons with scattering rate of $\tau_b$ between time t and t+$\Delta$t is calculated as, $\textrm{P}_{scat}=1-\exp(\frac{-\Delta t}{\tau})$.  If the probability of scattering, P$_{scat}$, is greater than a chosen random number R$_{scat}$, then the scattering is occurred. As a phonon scatters, its frequency, branch and direction will be re-sampled from the cumulative density function \cite{Shomali2017}. Here, the time step, $\Delta$t is chosen to be smaller than the minimum scattering rate of the phonons which are sampled during the simulation. To make sure that the rate of formation of phonons of a certain state is equal to its rate of destruction, the distribution function after scattering, F$_{scat}$, has to be modified by the probability of scattering, $\textrm{F}_{scat}(\tilde{T})=\frac{\sum_{j=1}^{i}\textrm{N}_j(\tilde{T}) \ \times \textrm{P}_{scat,j}}{\sum_{j=1}^{N_{b}}\textrm{N}_j(\tilde{T}) \ \times \textrm{P}_{scat,j}}$. The whole simulation procedure is more fully discussed in \cite{Mittal2011,2Shomali2017}.

\section{Numerical method}
\label{Numerical method}
	
For finding the time step, the whole frequency range of each branch of LA, TA, and ZA, is divided into 1000 intervals. For every branch, the scattering rates are calculated for the phonons with the all possible 1000 frequencies in that branch. Then the phonon relaxation time is computed by reversing the scattering rate. Also, the velocities of the all phonons with 1000 different frequencies are calculated for LA, TA, and ZA dispersion relation curves. Dividing the mesh size by the velocities, the phonon traveling time is obtained. The minimum value of these acquired times is taken to be the time step. As a consequence, the time steps of 3.47$\times$10$^{-13}$, 2.07$\times$10$^{-13}$, 3.46$\times$10$^{-13}$, 2.26$\times$10$^{-13}$, 2.57$\times$10$^{-13}$ are found, respectively, for graphene, blue phosphorene, germanene, silicene, and the MoS$_{2}$. Mesh-independent test has also been performed to investigate the numerical convergence of the results of the all simulated channels. It is seen that using a uniform mesh size of 100$\times$100 in XY-plane is appropriate to generate the mesh independent plots during the whole time of the simulation. Also, in the cube of 10$\times$100$\times$0.27 nm$^{3}$, located in the center of the all low-dimensional materials of graphene, blue phosphorene, silicene, germanene and MoS$_{2}$, the heat is generated due to the electric current transport. This heat generation zone should be incorporated in the Boltzmann equation. This term is simulated as a resource that heats up the desired zone by releasing the phonons into it. The procedure is completely explained in \cite{Wong2011,Wong2014}.

\section{Results and Discussions}
\label{results}

\begin{figure}
\vspace*{-2.53cm}
\centering
\hspace*{-0.85cm}
\vspace*{-5.7cm}
\includegraphics[scale=0.38]{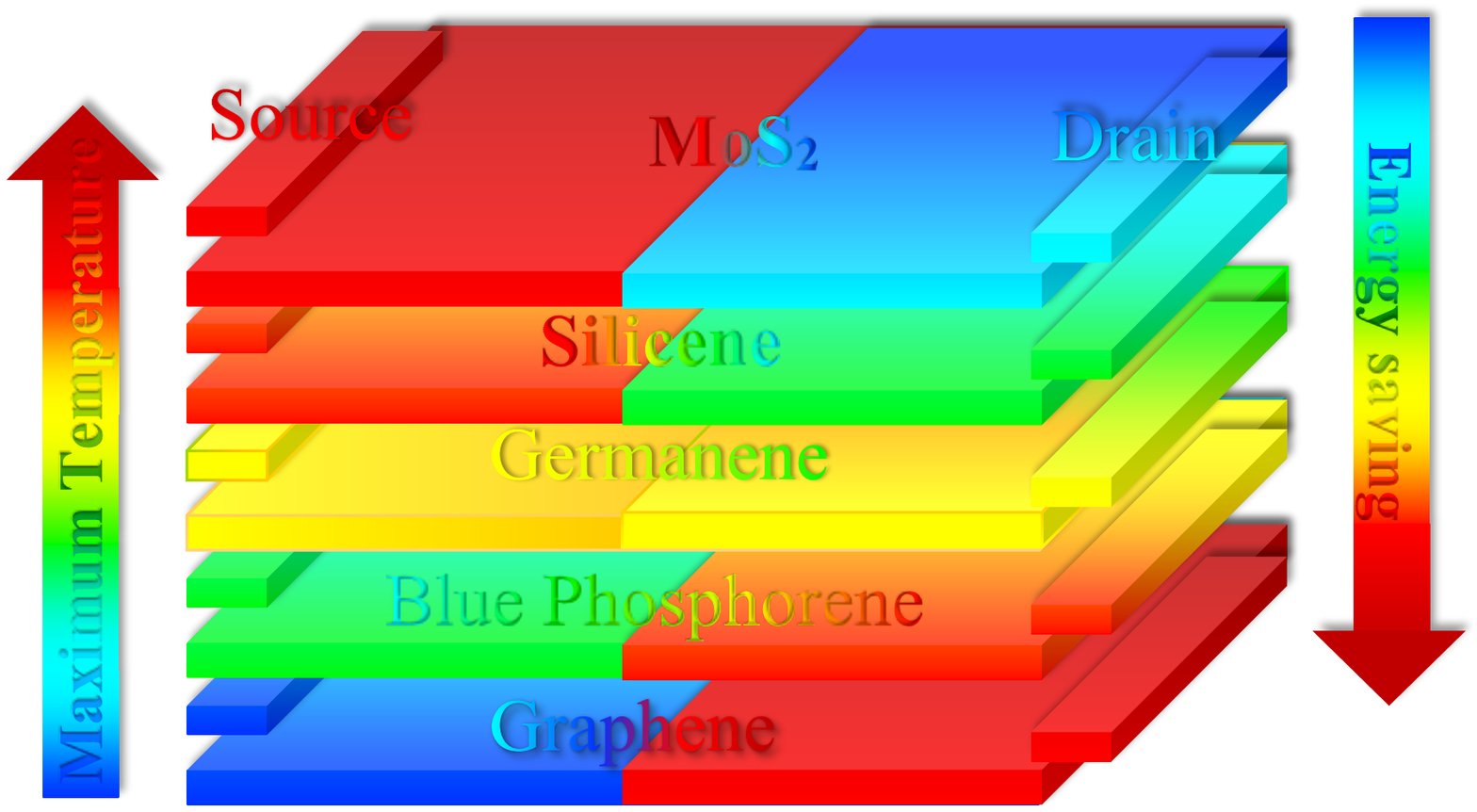}
\caption{\label{conclusion} Graphene, blue phosphorene, germanene, silicene, MoS$_2$ as the transistor channels show the least up to maximum peak temperature rise which suggest them reversely as the maximum to minimum thermal efficient channels concerning the nano energy saving of field effect transistors.}
\end{figure}

For discovering the best energy efficient systems, the thermal behavior such as the heat spreading and the maximum temperature reached due to the heat generation, in new proposed low-dimensional MOSFETs are studied. The five studied nano-systems are affected by the self-generated heat (mostly come from the Joule-Heating process) during the first 200 ps of the computation time. For the second 200 ps, the low-dimensional nano-systems are cooled. The result of this work can be summarized in the Fig. \ref{conclusion}. The five studied 2D crystalline materials of graphene, blue phosphorene, germanene, silicene and MoS$_2$ as the transistor channels, show subsequently, the least up to maximum peak temperature rise which suggest them reversely as the maximum to minimum thermal efficient channels concerning the nano energy saving of the field effect transistors. Although, the graphene sheet, thermally perfect for silicon substitution, does have many electrically difficulties, blue phosphrene in addition to other advantages is also electrically and thermally suitable for nano-electronic industry. The highlighted thermal advantages highly recommend the blue phosphorene, as a very promising high-performance low-dimensional material for next-generation nano-devices. In the following, more results are given to clarify the main finding of the research.

\begin{figure}
\vspace*{-0.04cm}
\centering
\hspace*{-0.6cm}
\vspace*{-0.78cm}
\includegraphics[scale=0.42]{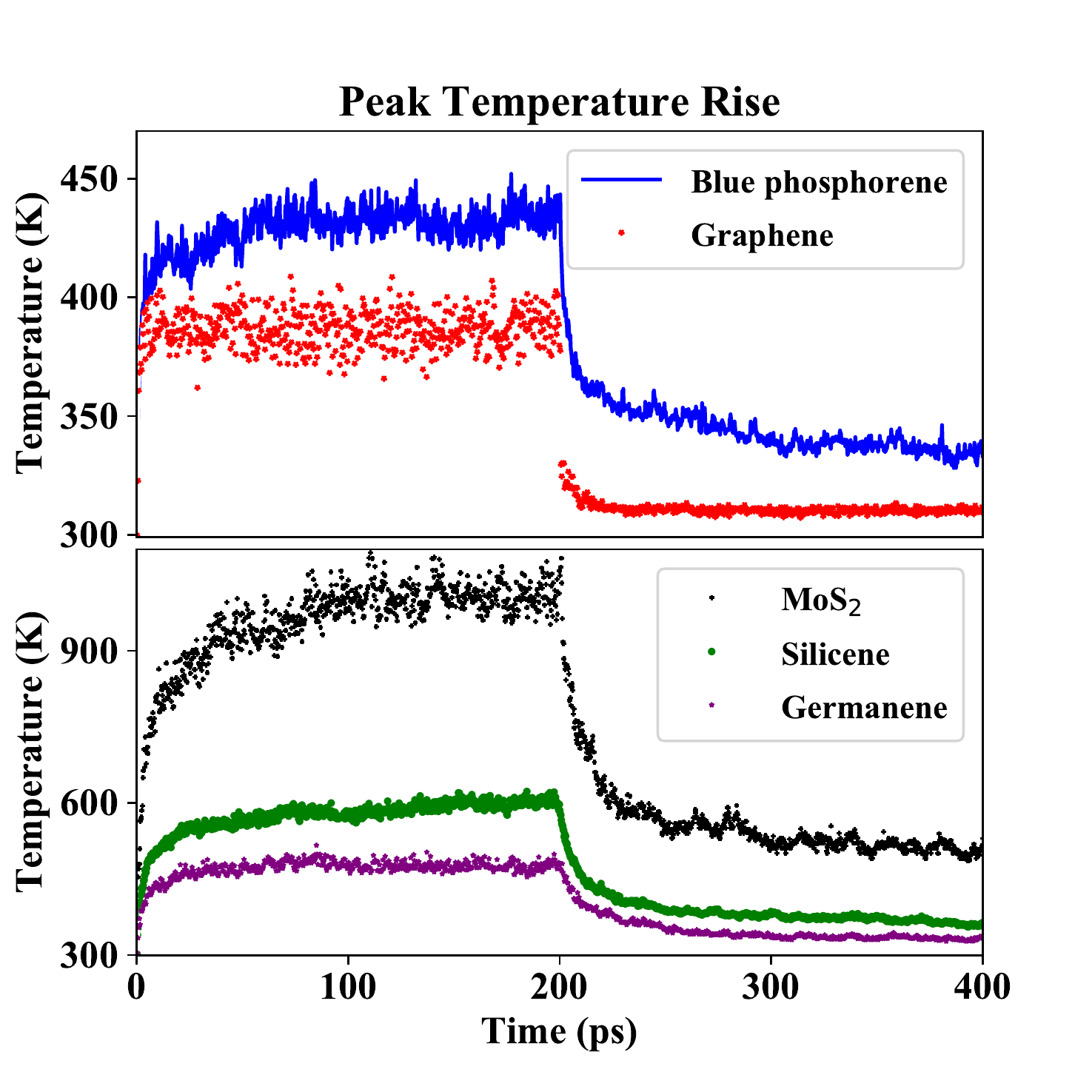}
\caption{\label{1} The peak temperature rise versus time at x,y plane when z=z$_{max}$ and t=200 ps for (a) graphene and blue phosphorene, (b) germanene, silicene, MoS$_{2}$ low-dimensional materials. Graphene and blue phosphorene do have obviously the least temperature rise which suggest them as most thermally efficient and low consumption materials.}
\end{figure}

As the reliability of the transistors depends on the maximum temperature it makes, the peak temperature rise versus the passed time is thoroughly explored. The Figs. \ref{1}(a) and (b) present, subsequently, the maximum temperature behavior for graphene/blue phosphorene, and also germanene/silicene/MoS$_{2}$ channel transistors. As seen, the peak temperature trends are similar for all the five materials. Further, while the graphene and blue phosphorene plots are more common in low values of maximum temperature, the germanene, silicene and MoS$_{2}$ follow the worse situation. Much less maximum temperature of the transistors including the graphene or blue phosphorene guarantees the much less energy needed for cooling down these nano-devices to the temperature which they can operate on. This finding confirms the more suitable nano energy saving conditions of blue phosphorene and graphene for the usage in nano-electronic devices technology which is preciously indicated in Fig. \ref{conclusion}.

\begin{figure}
\vspace*{-0.3cm}
\centering
\hspace*{-0.9cm}
\vspace*{-0.3cm}
\includegraphics[scale=0.42]{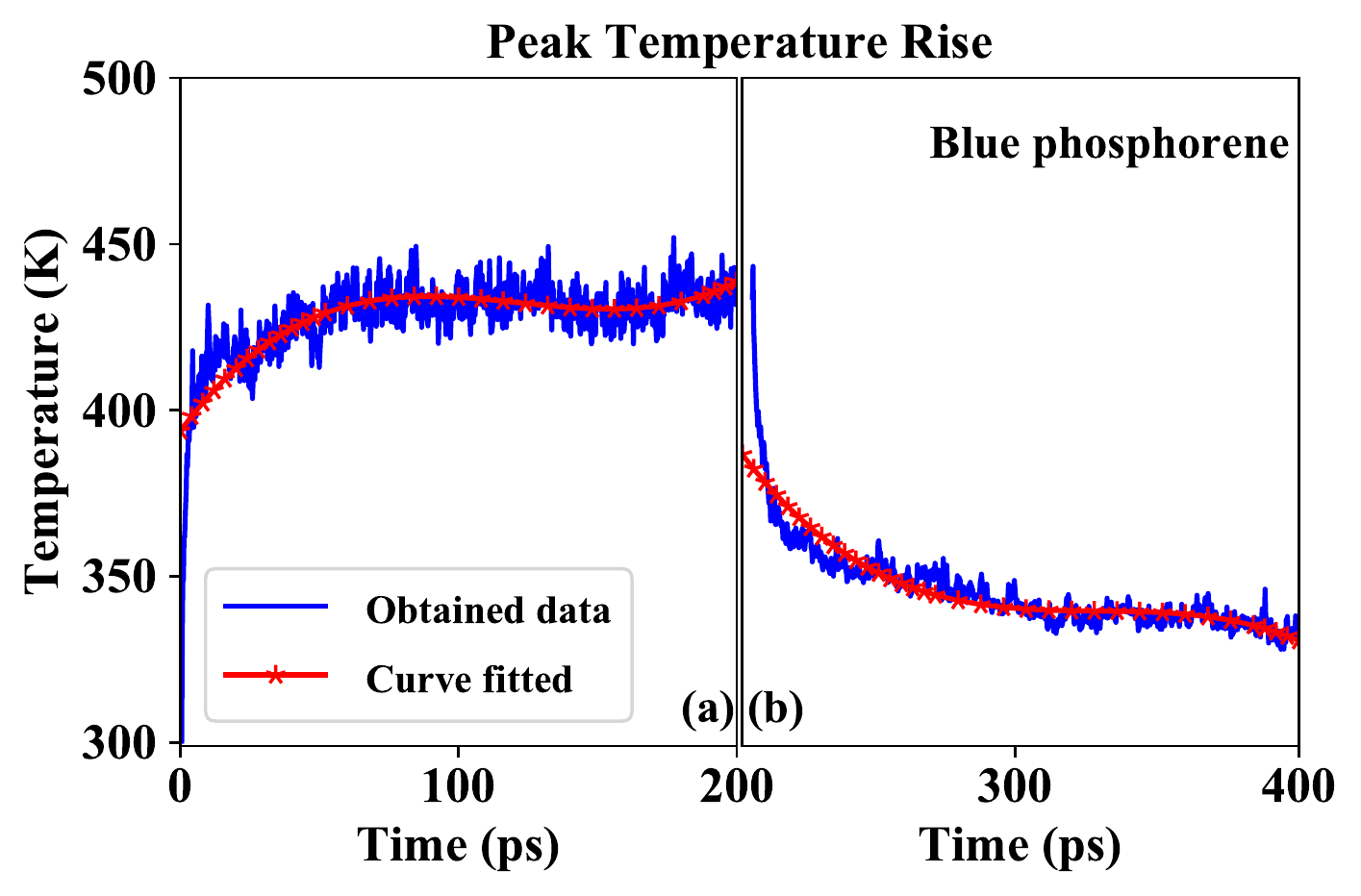}
\caption{\label{2} The obtained data of maximum temperature and the three order polynomial curves fitted to it, presenting the existence of the temperature jumps and the local minimum/maximum temperature when z=z$_{\rm max}$ at (a) t=200 ps and (b) t=400 ps.}
\end{figure}

In spite of the usual increasing behavior of the highest temperature during the presence of the heat generation, every plot of the low-dimensional materials presents the local minimum or maximum temperature and also the existence of the temperature jump. Such behavior is more precisely illustrated for blue phosphorene in Fig. \ref{2}(a-b). As seen, the three order polynomials of the form T=$\it a$t+$\it b$t$^2$+$\it c$t$^3$+$\it d$ with $\it a$=1.128, -13.35; $\it b$=1.0$\times$10$^{-2}$, 4.37$\times$10${^-2}$; $\it c$=2.75$\times$10$^{-5}$; 4.72$\times$10$^{-5}$ and $\it d$=393.86, 1684.8 are fitted to the original maximum temperature-time plot. Through looking thoroughly, local maximum at t=84 ps, the local minimum at t=160, the absolute maximum at t=200 ps and two temperature jumps in t=250, and 390 ps are clearly recognizable. The local minimum/maximum temperature and the temperature jump manners attribute to the change of the dominant phonons participated in heat transfer from slow-ZA branch phonons to the fast LA/TA phonons. The Table~\ref{Tabpercentage} demonstrates the analysis, related to the amount of the contribution of the phonons of each branch in heat transport. As seen, for example for blue phosphorene, when t=84 ps, most of the committed phonons belong to the flexural direction ZA branch which has very low sound speed. Hence, the heat dissipation decreases and consequently the energy inside the hot region increases which itself causes the generation of more phonons with high frequency. These new produced phonons make the temperature to rise more rapidly. Inevitably, the temperature jump takes places. Therefore, the high predominance of ZA phonons over LA/TA phonons is the reason for all temperature jumps occurring during the heating and cooling the transistor. Conversely, the usual attitude of the peak temperature rise is also disturbed owing to the occurrence of local minimum. This posture, in reverse, is ascribed to the dominance of the LA phonons over the ZA ones. As seen in Fig. \ref{2}(a), at t=160 the percentage of the LA phonons taking part in heat transfer is 56.96$\%$ while only 37.90$\%$ of the total phonons are the ZA participated phonons. Faster phonons disperse more heat and so despite the heat generation zone throw in more heat, they reduce the temperature temporarily.
\begin{table}[htbp]
\caption{The percentage of the number of the phonons of the LA, TA, and ZA branches which contribute to the heat transfer for the material blue phosphorene.}
 \label{Tabpercentage}
\begin{small}
 \centering
\hspace*{3.2cm}
  \begin{tabular}{|c|c|c|c|}
  \hline
t (ps) &  LA(\%)  &
TA(\%) & ZA(\%) \\
 \hline
 84  & 18.47 & 8.92 & 69.05  \\
 \hline
160 & 56.96 &  5.13 & 37.90\\
 \hline
200 & 33.85 & 9.63 & 56.51\\
\hline
250 & 8.15 & 25.98 & 65.87 \\
 \hline
390 & 10.44 & 22.15 & 67.40\\
 \hline
\end{tabular}
\end{small}
\end{table}

Regardless of the fluctuations and the local minimum/maximum, the peak temperature somehow presents monotonous and homogeneous behavior over the time t=10 ps. As previously indicated, the low-dimensional systems are three dimensional channels with the thickness of the order of angstrom which are considered to exchange energy with the environment from the bottom boundary. At t=0, when the transistor is switched on, the heat starts to be generated inside the heating zone. This strongly increases the temperature of the zone up-to time t=10 ps. As the thickness of these materials is amazingly short, the heat reaches the bottom boundary in 10 ps and the cooling process, immediately after turning on the heating zone, begins. Ergo, although, the heat is produced in the hot zone, the heat also dissipates from the boundary and that is the reason why the temperature does not grow unlimitedly.

\begin{figure*}
\centering
\includegraphics[width=0.75\textwidth]{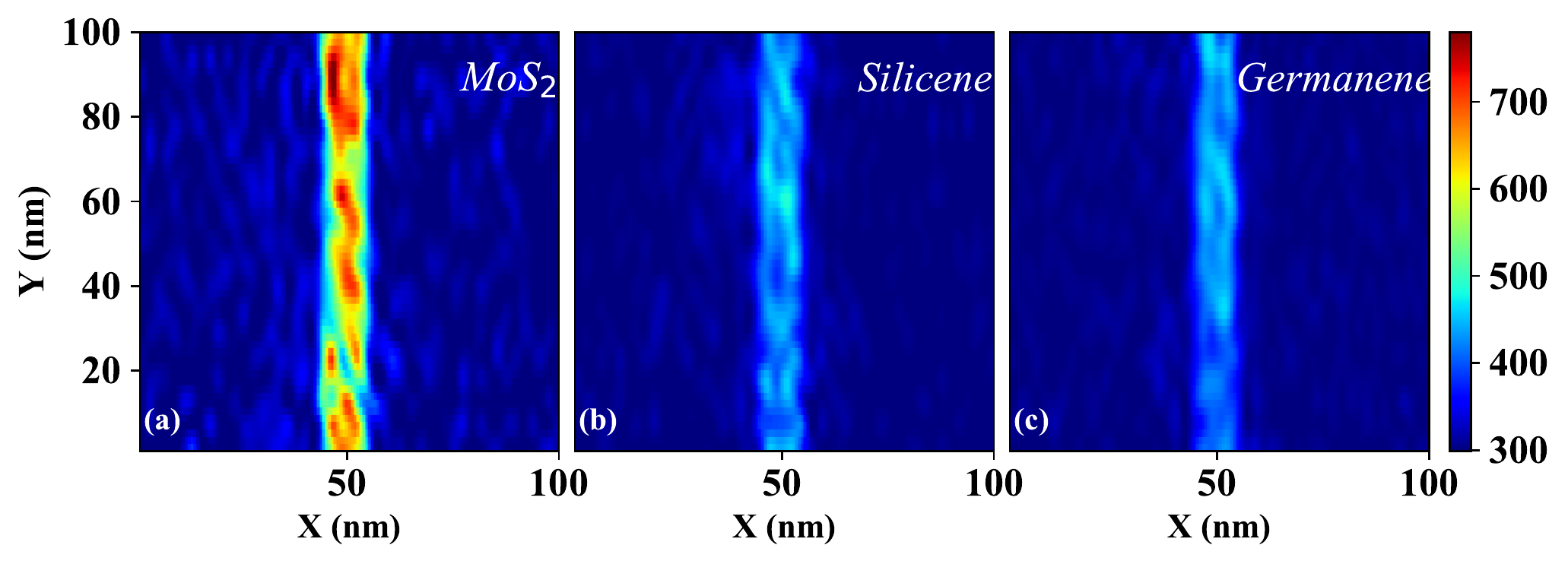}
\caption{\label{sghotspot2} The temperature profile at XY plane and the generated hot spot at t=200 ps for (a) the MoS$_{2}$, (b) the silicene, and (c) the germanene, confirming that the MoS$_{2}$ renders very poor thermal conditions due it is much more higher peak temperature rise relative to the other investigated low-dimensional materials.}
\end{figure*}

\begin{figure}
\centering
\includegraphics[scale=0.42]{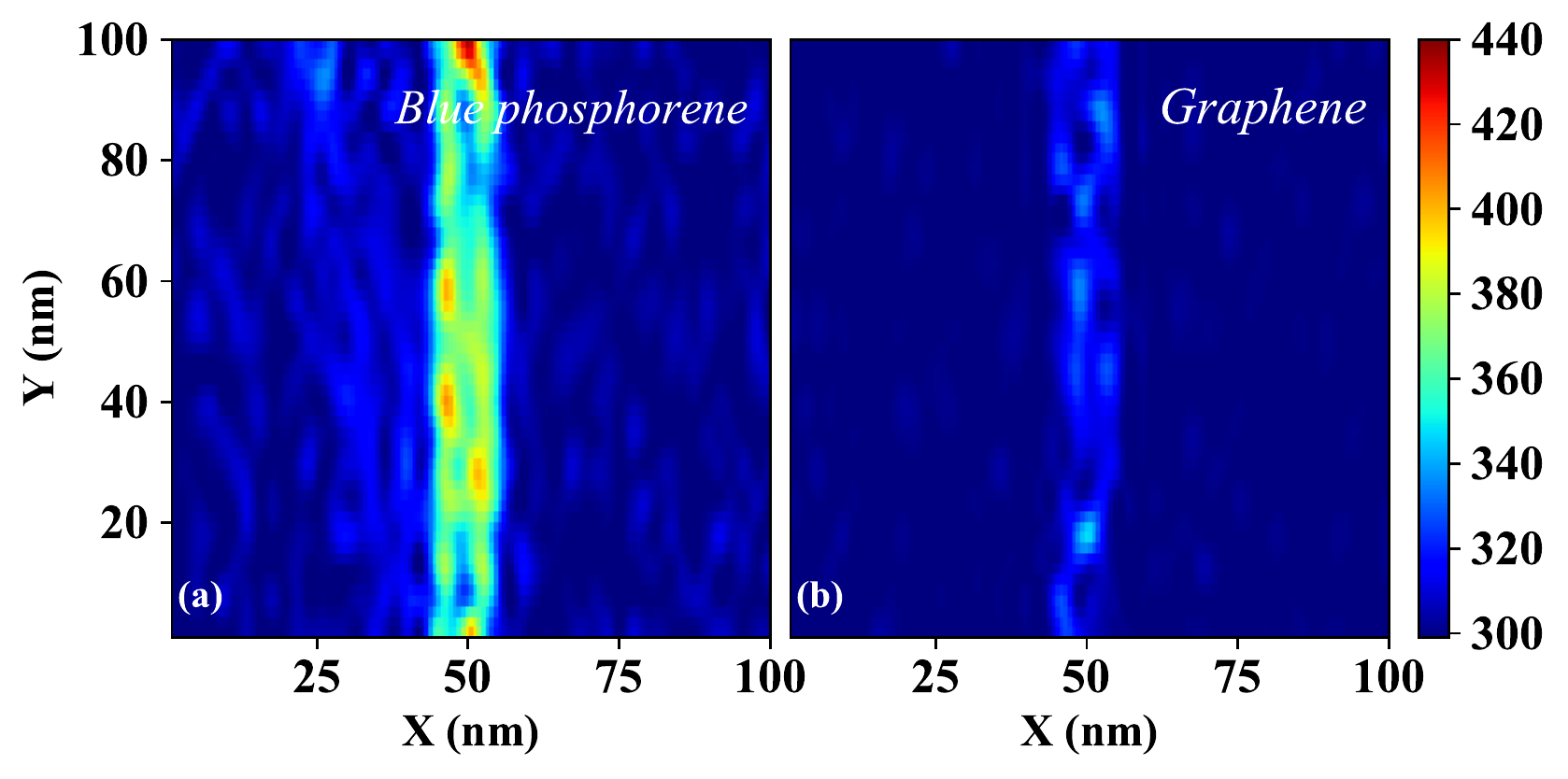}
\caption{\label{bpghotspot} The colormap demonstrating the hot spot condition for the two studied channels of (a) the blue phosphorene and (b) the graphene stabilizing these two materials as the very suitable candidates for fulfilling the thermal management of the new generation field effect transistors. Specifically, blue phosphorene is highly more recommended due to its simultaneous electrically benefits.}
\end{figure}

Another finding is the formation of the hotspots in every five channels. When the heat generation zone is switched on, due to the injection of the hot phonons to the transistor, the overall energy of the MOSFET specially near the heated zone got increased. At this time the most of the phonons taking part in the heat transport are the ones in flexural direction (ZA mode). These phonons are much slower than the phonons in longitudinal or transverse modes transferring through the transistor. Hence, the nano-device becomes hot while most of the phonons presented in hot region are ZA phonons. Consequently, these slow phonons do not carry the heat from the hot zone to the surrounding places and the heat falls into the trap and accordingly the hot spot is formed. Figs. \ref{sghotspot2}(a-c) and \ref{bpghotspot}(a-b) exhibit the created hotspots in five investigated low-dimensional materials at t=200 ps just as the time that the heat generation zone is turned off. As the figures suggest, the MoS$_{2}$, silicene, germanene, blue phosphorene, and graphene, subsequently, does have the hottest hotspot. Put differently, these plots confirm that due to the high temperature of the hotspots of the MoS$_{2}$, silicene, germanene, cooling down these substances are more difficult than that of the graphene and blue phosphorene. As mentioned before in Fig. \ref{conclusion}, the obtained behavior of the hotspots suggest these two substances as the most appropriate low-dimensional materials for the most efficient energy saving/consumption of the transistors.

In addition, as the heat source is switched off, the hotspots of low-dimensional materials start to lose heat. During the whole 200 ps of cooling process, the ZA phonons are the dominant phonons taking part in heat dissipation. At some times such as t=275 and 334 ps, the percentage of the predominant phonons increases notably. As discussed before, the more slow phonons make the heat dissipation take place gradually which results in the union of the phonons and appearance of the temperature jump. Fig. \ref{xy,z=kmax;t=400} (a-e) present the steady state behaviors of the all five investigated materials at t=400 ps. As the figures recommend, the graphene, blue phosphorene, germanene, silicene and MoS$_{2}$, subsequently, are more likely to cool faster. That is to say, the first two channels represent the hotspots with the least value temperature and therefore after 200 ps of going out the heat source, they are superlatively better chilled. This finding confirms the statement presented in the Fig. \ref{conclusion} emphasizing the much less energy consumption of the graphene and the blue phosphorene.

Moreover, the heating and cooling procedures of the formed hotspots of the blue phosphorene and the silicene are looked thoroughly. Both materials are started to be heated at t=0 ps. As seen in Figs. \ref{bp,si}(a), immediately after switching on the heating zone at t=2 ps, the hotspots are created in both systems. As the Figs. \ref{bp,si}(b-d) present, the hotspots in silicene get more hotter and localized than the one formed in blue phosphorene, evolving with time. The phrase more localized means that the heat is more likely to concentrate in the heating zone. In other word, after passing the same time from turning on the heating source, the phonons has started to dissipate the heat through the blue phosphorene device, while the heat is still trapped in the silicene channel. It is obvious that as the time passes, the flux of heat in two materials start to spread all over the nano-system. The Figs. \ref{bp,si}(e-g) present that in the trend of evolution of the temperature distribution and going towards to the steady-state, the blue phosphorene is always ahead of germanene, silicene and MoS$_{2}$. Also, it is obtained that the steady-state temperature of the blue phosphorene is much more less than that of other materials except the graphene (see Fig. \ref{bp,si}(g)).

Finally, considering all the results obtained simulating the low-dimensional field effect transistor, we come back to the first announced proclamation in Fig. \ref{conclusion} that the blue phosphorene and graphene are overwhelmingly much more thermally suitable for transistor technology. 
\section{Conclusions}
\label{con}

With the progression in the integrated circuits (IC) design technology, the number of the fabricated FETs in a die has now reached more than a million. This size reduction cannot continue forever due to limitation existed in miniaturized silicon nano-channels such as the huge growth in energy consumption and the heat dissipated due to the leakage current. As a consequence, finding the best silicon substitution for the efficient energy usage is a vital and noteworthy need for MOSFET technology. Two-dimensional crystalline materials owing to their zero leakage current and the extraordinary thermal and electrical properties, are the adequate substances for use as the channels of the field effect transistor. While numerous elucubrations have been performed concerning the electrical advantages and disadvantages of different two-dimensional channels, a few works have studied the thermal behaviors of such systems. Complete knowledge of the thermal behaviors of the MOSFET channels is inevitable since the self-heating and the method of cooling the transistors determine the reliability of these nano-devices. In this paper, five  low-dimensional materials including graphene, blue phosphorene, germanene, silicene and MoS$_{2}$ are studied from the non-Fourier thermal point of view by using the Phonon Monte-Carlo simulation. In particular, the maximum temperature, the indication of the reliability of the nano-device in nano-electronics, reached by these materials under the self-heating phenomena is calculated. The analysis reveals that the graphene and the blue phosphorene have the least peak temperature rise between the all five inquired low-dimensional nano-systems at every time during the simulation. In other words, for these materials, the formed hotspot is cooler in comparison to that of the others. The less self-heated channels, are more likely to be the silicon channel replacement due to their easier cooling process and consequently not reaching the temperature threshold level. Moreover, after switching off the heat generation zone, these two materials attain the steady state faster with lower temperature. These results, together with the electrical inefficiency of the graphene sheet, nominates the blue phosphorene as the best embedment choice of the old-world silicon channel in the next generation of the field effect transistors.

\newpage
\newpage
\newpage

\begin{figure*}
\centering
\includegraphics[scale=0.52]{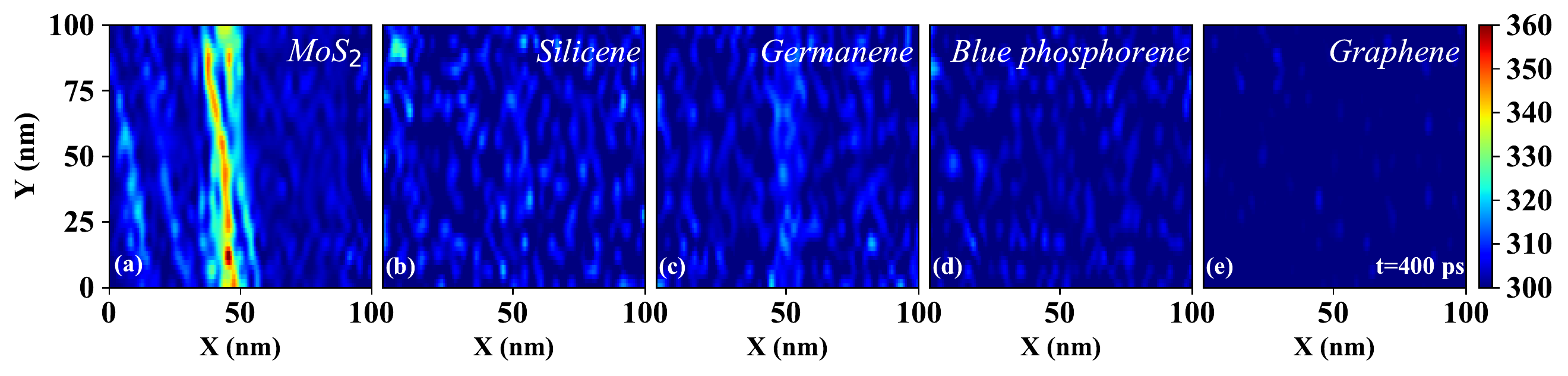}
\caption{\label{xy,z=kmax;t=400} (a-e) The temperature distribution in XY plane 200 ps after turning off the heat generating zone at t=400 ps for MoS$_{2}$, silicene, germanene, blue phosphorene, and the graphene, respectively, declaring the much slower move towards to the steady state with the higher steady state temperature in the order given.}
\end{figure*}

\begin{figure*}
\vspace*{-1.2cm}
\centering
\hspace*{-1.1cm}
\vspace*{-0.43cm}
\includegraphics[scale=0.56]{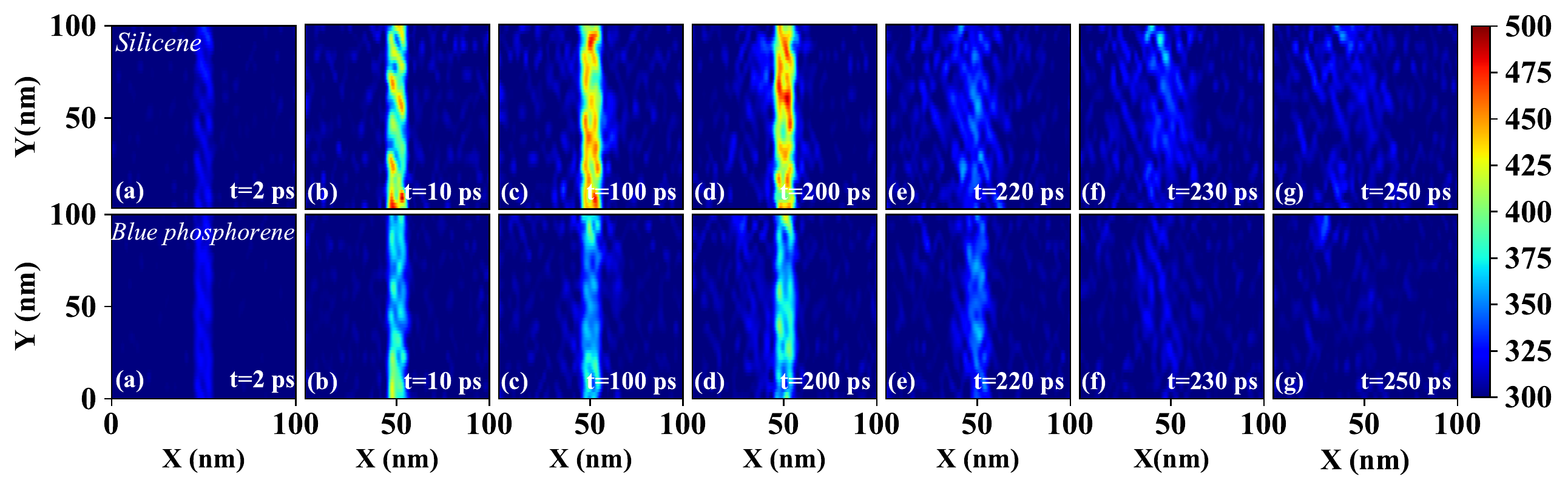}
\caption{\label{bp,si} The condition of the hotspot during the simulation in XY plane at (a) t=2 ps, (b) t=10 ps, (c) t=100 ps, (d) t=200 ps, (e) t=220 ps, (f) t=230 ps, and (g) t=250 ps. The upper and lower plots, respectively, belong to the silicene and blue phosphorene channels which presenting the immediate formation of the hotspots, always smaller in value and relaxing in lower temperature at steady state for the blue phosphorene.}
\end{figure*}

\end{document}